\documentclass[twocolumn,showpacs,amsmath,amssymb,prd,nofootinbib]
{revtex4}
\newcommand{\beq}{\begin{equation}}
\newcommand{\eeq}{\end{equation}}
\def\eeqno#1{\label{#1}\end{equation}}
\def\sss{\scriptscriptstyle}
\def\^#1{^{\sss #1}}
\def\_#1{_{\sss #1}}

\def\az{a_0}
\def\msun{{\rm M}_{\odot}}
\def\kms{~{\rm km~ s}^{-1}~}

\def\kpc{~{\rm kpc}}
\def\mpc{~{\rm Mpc}}

\def\cmss{{\rm cm~s^{-2}}}
\def\m{\mu}
\def\C{\Gamma}
\begin{document}

\title{MOND and the unique void galaxy KK246}
\author{Mordehai Milgrom}\affiliation{DPPA, Weizmann Institute of
Science, Rehovot 76100, Israel}
\begin{abstract}
MOND predictions are compared with the mass discrepancy, $\C$ (the dynamical-to-baryon mass ratio) deduced from the recently measured rotation curve, for the gas-rich, dwarf galaxy KK246, ``the only galaxy observed in the local void''. KK246 is special in at least two regards: a. It is, to my knowledge, the record holder for the largest mass discrepancy deduced from a rotation curve, $\C\approx 15$.
b. It is very isolated, residing in a large, very empty void.
I also discuss another extreme case: Andromeda IV, a dwarf considered here for the first time in light of MOND, with a very large mass discrepancy, $\C\approx 12$, also conforming accurately to the MOND prediction. In both cases, MOND predicts $\C$, or the total dynamical mass at the last observed radius, from only the knowledge of the small mass of baryons.
If MOND is accepted as the root of the mass discrepancy, these are just two more expected, albeit reassuring, conformities. However, in the framework of the dark-matter paradigm--where the mass discrepancy is strongly dependent on the buildup history of a galaxy--every new such conformity with a tight law is another difficult-to-understand surprise, and does carry a new import: What, in the $\Lambda$CDM paradigm, would prevent such galactic baryons from residing in a halo of half, or twice, the observed rotational velocities, instead of selecting exactly the velocities predicted by MOND? This conundrum is especially poignant for KK246, whose great isolation points to a relatively unique buildup history. This note underscores the individual importance of each galaxy as a new test, as opposed to the view of them all as a statistical ensemble.

\end{abstract}

\maketitle
\section{The mass discrepancy in KK246}
Kreckel \& al. (2011) have recently presented their measurement of the HI distribution, and the rotation curve, of the gas-rich, low-surface-density galaxy, KK246. It is said to be the only confirmed galaxy in the local (Tully) void. It boasts ``an extremely extended HI disk'', viable, presumably, because of the extreme isolation of the galaxy: It has no companions discovered within $3\mpc$ of itself. The Tully void is very large and empty. Its emptiness, and that of other voids, has been brought up as a challenge to the $\Lambda$CDM paradigm, which predicts much more populous voids (e.g., Tikhonov \& Klypin 2009, and references therein). But here I consider another issue that poses a serious challenge to the $\Lambda$CDM paradigm, relating to the internal dynamics of KK246.
\par
Kreckel \& al. (2011) quote the following characteristics of KK246, normalized to their adopted distance of $D_a=7.83\mpc$: A total B-band luminosity $L_B=4.6\cdot10^7 (D/D_a)^2L_\odot$, HI mass of $M\_{HI}=1.05\cdot10^8 (D/D_a)^2\msun$, corrected to give a gas mass of $M\_{gas}\approx 1.5\cdot10^8 (D/D_a)^2\msun$. They adopt a stellar mass of $M_*=5\cdot10^7 (D/D_a)^2\msun$, which corresponds to $M_*/L_B\approx 1(M/L)_{\odot}$. The deduced rotation curve rises slowly, and roughly levels off at $\approx 42\kms$, beyond galactic radius of $\sim 5\kpc (D/D_a)$. The last measured radius is $r_m\approx 7(D/D_a)\kpc$. The dynamical mass within $r_m$ is then $M_{dyn}(r_m)=G^{-1}V^2(r_m)r_m\approx 2.9 \cdot10^9 (D/D_a)\msun$. This give an observed mass discrepancy (dynamical-to-baryonic mass ratio) at $r_m$ of
 \beq\C(r_m)\approx 14.5(D/D_a)^{-1},\eeqno{aom}
for the above $M/L$ value. This is somewhat at odds with the value of $\C(r_m)\approx 20(D/D_a)^{-1}$ quoted by Kreckel \& al., who give a value of $M_{dyn}(r_m)=\approx 4.1 \cdot10^9 (D/D_a)\msun$. This seems to be traceable to their using in the expression for $M_{dyn}$ the velocity gotten as $W_{20,i}/2$, the inclination-corrected half line width at 20\% intensity, instead of using the more appropriate $V(r_m)$.
\par
Now consider the value of $\C(r_m)$ predicted by MOND. The {\it measured} centripetal acceleration at $r_m$ is $g=V^2(r_m)/r_m\approx 8\cdot 10^{-10}(D/D_a)^{-1}\cmss\approx 0.068\az(D/D_a)^{-1}$, where $\az$ is the MOND constant, taken
as $\az=1.2\cdot 10^{-8}\cmss$. MOND predicts $\C(r_m)=1/\m(g/\az)$, where $\m(x)$ is the appropriate MOND interpolating function. Because here $g/\az\ll 1$, the dependence on the exact form of $\m(x)$ is very weak, as in this limit $\m(x)\approx x$. For example, using this last approximation, MOND predicts [note the different $D$ dependence from eq.(\ref{aom}); MOND can be used to determine distances]
 \beq\C(r_m)\approx14.7(D/D_a), \eeqno{bom}
while if for $x\ll 1$ a better approximation is $\m(x)\approx x/(1+x)$, which seems to be preferred by rotation-curve analysis, then MOND predicts
 \beq \C(r_m)\approx 15.7(D/D_a).  \eeqno{gom}
The above prediction is based on the {\it measured} centripetal acceleration. We can also start from the total baryonic mass (which is rather well within $r_m$), which for $M_*/L_B= 1(M/L)_{\odot}$ is $M_{bar}\approx 2\cdot 10^8(D/Da)^2\msun$. This gives a Newtonian acceleration at $r_m$ of $g\_N\approx 4.7\cdot 10^{-3}\az$. This, in turn, gives a MOND acceleration $g/\az\approx
(g\_N/\az)^{1/2}\approx 0.069$ (note the different dependence on the distance of this acceleration from the one deduced from the rotation curve; the two agree very well if $D\approx D_a$). This then predicts a rotational speed at $r_m$ of $42(D/D_a)^{1/2}\kms$, just as observed. So, in effect, MOND uses only the small amount of baryonic matter to predict the much larger, apparent mass of ``dark matter'', which within $r_m$ is $M_{DM}\approx 14M_{bar}$.
\par
We need to keep in mind several sources of uncertainty: (i)  Kreckel \& al. mention the possible effect of non-circular motions that might cause some error in determining the acceleration from the rotation curve (ii) The stellar $M/L$ value might differ from that adopted, but because the stars contribute only about 0.25 of the baryonic mass for the adopted value, the error introduced by this will not be large. (iii) Asymmetric-drift corrections are not expected to be important for galaxies of this type (see, e.g., estimates of the effect in Begum \& Chengalur 2005), but could cause changes in the velocities of a few $\kms$. For these reasons, and also because the distribution of the stellar mass in KK246 is not known (it is said to be ``optically quite irregular, with no clear nuclear or disc structure''), I have not attempted to calculate the MOND rotation curve. But, it should be noted that the observed curve looks very much like the MOND curve for a low-acceleration exponential disc with a scale length of $\sim 1.5 \kpc$ (e.g., Milgrom 1983), which this galaxy is, approximately.

\section{Andromeda IV}
And IV is another extreme case. Chengalur \& al. (2008) find an extreme $M_{dyn}/L_b=237(D/D_a)^{-1}$, at their $r_m\approx 6.25(D/D_a)\kpc$, with the adopted distance of $D_a=6.11 \mpc$.\footnote{The distance to And IV is somewhat uncertain. In fact, MOND, if accepted, would give the best distance estimate for this galaxy.} They give $M\_{HI}\approx 1.8\cdot 10^8(D/D_a)^{2}\msun$, and so,  $M_{gas}\approx 2.5\cdot 10^8(D/D_a)^2\msun$.
They also determined the rotation curve, which rises gently and flattens at $V\_{\infty}\approx 46 \kms$, beyond $4(D/D_a) \kpc$, and up to $r_m$.
The mass distribution itself is not available to me; so a MOND rotation curve cannot be calculated. And IV also has an extreme value of   $M\_{HI}/L_b=13$; so  $M\_{gas}/L_b\approx 18$, and thus stars hardly contribute to the baryonic mass, practically obviating uncertainties due to the stellar $M/L$ value. Taking a nominal value $M_*/L_B= 1(M/L)_{\odot}$, with $L_B\approx 1.4\cdot 10^7(D/D_a)^2 L\_{\odot}$, the baryonic mass (well within $r_m$) is $M\_{bar}\approx 2.64\cdot 10^8(D/D_a)^2\msun$. The measured centripetal acceleration at $r_m$ is $g=V^2(r_m)/r_m\approx 1.1\cdot 10^{-9}(D/D_a)^{-1}\cmss\approx 0.09\az(D/D_a)^{-1}$. For different choices of $\m(x)$, as above, this corresponds to a MOND predicted
 \beq \C(r_m)\approx (11-12)(D/D_a).\eeqno{numa}
 The predicted dynamical mass at $r_m$ is thus
$M_{dyn,mond}\approx (2.9-3.2)\cdot 10^9(D/D_a)^3\msun$. This is to be compared with the observed dynamical mass of $M_{dyn,measured}\approx 3.1\cdot 10^9(D/D_a)\msun$, a very good agreement for $D\approx D_a$.
\section{Discussion}
For both galaxies, $r_m$ is already in the regime where the rotation curve is approximately flat. Thus, the above comparisons amount to checking that the MOND predictions of the mass-asymptotic-velocity relation (aka, the baryonic Tully-Fisher relation) is confirmed. Indeed, both galaxies accurately satisfy this relation: From the baryonic mass of KK246 [with the adopted distance, and $M^*/L_B=1(M/L)\_{\odot}$], MOND predicts a flat velocity of $V\_{\infty}=(M_{bar}G\az)^{1/4}\approx 42(D/D_a)^{1/2}\kms$, as observed. Similarly, in And IV, MOND predicts from the baryonic mass $V\_{\infty}\approx 45.3(D/D_a)^{1/2}\kms$, again, very near what is observed.
So, these galaxies would fall right on the line predicted by MOND for the baryonic Tully-Fisher relation, as in McGaugh (2011) (whose analysis does not include the present galaxies). 
\par
We would be amiss, however, to consider the evidence from these galaxies as just two more points on the MOND $M_{bar}-V\_{\infty}$ line, with no particularly new import: If it is accepted
that MOND underlies the mass discrepancy, there isn't, indeed, much surprise in the present findings, as there isn't much surprise in finding that every newly discovered planetary system obeys Kepler's laws. However, for someone who believes that the mass discrepancy is a result of the detailed buildup of a galaxy, with DM and baryons each playing its own part, it should be a great surprise every time a new case is discovered, where the puny baryons seem to determine the dominant ``DM'', even in such extreme cases as the present ones.
\par
For example, Kreckel \& al. speak of ongoing gas accretion in KK246. If this is an important buildup mechanism, then it increases the baryonic mass, but not the DM mass. So the MOND prediction should have been correct only at one time.


This research was supported by a
center-of-excellence grant from the Israel Science Foundation. I am grateful to Stacy McGaugh and Bob Sanders for useful comments.

\end{document}